\documentclass[aps,amsmath,amssymb,amsfonts,twocolumn,prd,nofootinbib,longbibliography]{revtex4-1}
\pdfoutput=1

\usepackage{amsmath}    %
\usepackage{mathtools}  %
\usepackage{graphicx}   %
\usepackage{color}      %
\usepackage{hyperref}   %
\newcommand{\RefFig}[1]{{FIG.~\ref{#1}}}

\newcommand{\SoS}{c}

\newcommand{\stch}{^{\textrm{S}}}

\newcommand{\moch}{^{\textrm{M}}}

\newcommand{\RefEq}[1]{(\ref{#1})}

\newcommand{\RefSec}[1]{{Section~\ref{#1}}}

\newcommand{\RefSubsec}[1]{{Subsection~\ref{#1}}}

\newcommand{\RefStep}[1]{{Step~\ref{#1}}}

\begin{document}
	
	\title{Sound clocks and sonic relativity}
	\author{Scott L. Todd}
	\email{scott@todd.science}
	\affiliation{Centre for Quantum Computation and Communication Technology, School of Science, RMIT University, Melbourne, Victoria 3001, Australia}
	\author{Nicolas C. Menicucci}
	\email{ncmenicucci@gmail.com}
	\affiliation{Centre for Quantum Computation and Communication Technology, School of Science, RMIT University, Melbourne, Victoria 3001, Australia}
	
	\date{\today}

	\begin{abstract}
		Sound propagation within certain non-relativistic condensed matter models obeys a relativistic wave equation despite such systems admitting entirely non-relativistic descriptions. A natural question that arises upon consideration of this is, ``do devices exist that will experience the relativity in these systems?'' We describe a thought experiment in which `acoustic observers' possess devices called sound clocks that can be connected to form chains. Careful investigation shows that appropriately constructed chains of stationary and moving sound clocks are perceived by observers on the other chain as undergoing the relativistic phenomena of length contraction and time dilation by the Lorentz factor, $\gamma$, with $\SoS$ the speed of sound. Sound clocks within moving chains \textit{actually} tick less frequently than stationary ones and must be separated by a shorter distance than when stationary to satisfy simultaneity conditions. Stationary sound clocks \textit{appear} to be length contracted and time dilated to moving observers due to their misunderstanding of their own state of motion with respect to the laboratory. Observers restricted to using sound clocks describe a universe kinematically consistent with the theory of special relativity, despite the preferred frame of their universe in the laboratory. Such devices show promise in further probing analogue relativity models, for example in investigating phenomena that require careful consideration of the proper time elapsed for observers.
	\end{abstract}

	\maketitle

\section{Introduction}
\label{Sec: Introduction}

The repeated null results from experiments to detect the luminiferous aether towards the end of the 19\textsuperscript{th} century~--~most notably the null result of the Michelson-Morley experiment~\cite{Michelson1887}~--~culminated with many physicists, most notably George FitzGerald, Hendrik Lorentz, and Henri Poincar\'{e}, proposing mechanisms by which the aether was undetectable. FitzGerald~\cite{GERALD1889} and Lorentz~\cite{Lorentz1892a,Pauli1958} independently (1889 and 1892 respectively) suggested that objects contract parallel to their direction of motion. Woldemar Voigt~\cite{Voigt1887,Pauli1958} suggested modification of the time coordinate to ensure that the wave equation for light worked in all reference frames, and Lorentz~\cite{Lorentz1892,Pauli1958} also later introduced this same notion of `local time,' though unlike length contraction he did not assign any physical importance to it. Poincar\'{e}~\cite{Poincare1999}, however, realised a physical significance of this notion of local time as suggested by Lorentz in that it would be the time recorded on clocks synchronised using light signals.

Eventually, aether theory fell victim to Ockham's razor\footnote{Named in honour of William of Ockham, though often spelt ``Occam's razor''.}: Einstein's theory of special relativity sufficed to explain all of the same phenomena with the added simplification of not requiring an undetectable aether. It is important to realise that while the theory of special relativity won out over any of the aether theories, aether-based models still produce the exact same kinematic results as the theory of special relativity when treated correctly due to both theories exhibiting the exact same mathematical formalism~\cite{H.P.1949}.

In an effort to describe black holes, Unruh~\cite{Unruh2008} came up with an analogy in terms of sound waves propagating up a waterfall: if at some location along the waterfall the flow speed of the water exceeds the speed of sound in the water, it becomes impossible to sonically signal upstream anymore. This model is analogous to a black hole, except here sound takes the place of light. Such objects are called acoustic black holes or dumb holes (where `dumb' is a synonym for mute). It is worth noting that such a model possesses a preferred reference frame: the reference frame in which the water is stationary.

Under several assumptions (no gravitational back-reaction, an unquantised gravitational field, and that at the Planck scale the wave equation for quantum fields is still applicable), Stephen Hawking famously demonstrated that black holes may be expected to evaporate by radiating at a characteristic temperature~\cite{Hawking1974}. Whether the assumptions that Hawking made are valid is still unknown, and when Hawking's assumptions were scrutinised by Unruh~\cite{Unruh1976}, the result of black hole evaporation was initially put into doubt. Acoustic black holes possess comparative problems: at sufficiently small length scales the continuum description breaks down (the expected equivalent of the Planck scale for spacetime), the field fluctuations (phonons) interact with the background that they propagate on (in analogy to gravitational back-reaction), and the analogy to the gravitational field itself (the fluid flow) is unquantised.

Unruh, following the same process as Hawking, theorised that acoustic black holes should emit an acoustic analogue of Hawking radiation~\cite{Unruh1981}. Fortunately, in stark contrast to physics at the Planck scale, molecular physics, atomic physics, and fluid mechanics are well understood. While the validity of the underlying assumptions in Hawking's derivation of black hole radiation could not be directly tested, the comparative assumptions in the acoustic black hole model could be. Unruh~\cite{Unruh1995} and others (Brout \textit{et al.} (analytically)~\cite{Brout1995}, Corley (analytically)~\cite{Corley1997,Corley1998}, and Corley and Jacobson (numerically)~\cite{Corley1996,Corley1997a}) were able to show that acoustic black holes should radiate at a temperature as calculated using Hawking's approach for black holes even when the comparative assumptions in the acoustic black hole model were broken. This result demonstrates that the blackbody temperature of black holes as calculated by Hawking will not necessarily break down when the contributions of Planck scale physics are taken into account, providing us with some evidence that our current understanding of black hole thermodynamics could well be correct~\cite{Unruh2008}.

Models such as acoustic black holes are inherently nonrelativistic, yet nevertheless they appear to share some properties with relativistic systems~\cite{Barcelo2011a}. This class of models are referred to as \textit{analogue gravity models}. There are many examples of analogue gravity models. For example, quasiparticle production has been theorised~\cite{Fedichev2004} to occur in expanding Bose--Einstein condensates in analogy to particle production due to cosmic inflation. Bose--Einstein condensates have also been used to study analogue Hawking radiation, both theoretically~\cite{Barcelo2001a,Carusotto2008} and experimentally~\cite{Belgiorno2010,Weinfurtner2011}. An extensive list of the analogue gravity models known to exist up until 2011 can be found in the \textit{Living Reviews in Relativity} article by Barcel\'{o}, Liberati, and Visser~\cite{Barcelo2011a}.

It is natural to wonder how far these analogies can be pushed before they break down. How many of the features of general relativity appear in analogous form within these models? If all of the features of general relativity emerge in some analogue form, then under what assumptions does this occur? Is there a microscopic mechanism behind the analogue form of gravity, and can we use this knowledge to infer anything about the origins of gravitation in our universe? If it is not possible to make all of the features of general relativity emerge in some analogous form in these systems, then why do we find the emergence of some aspects of general relativity and not others?

In order to answer these questions it is useful to understand the experience of observers within analogue relativity systems since observers were crucial in the understanding of general relativity, specifically in explaining the physical interpretation of general covariance~\cite{Rovelli2004}. The principle of general covariance states that the laws of physics should be independent of choice of coordinate system, as coordinates do not exist in nature a priori. It can be shown, however, that smoothly dragging the gravitational field around on a background spacetime manifold (a mathematical process called an active diffeomorphism) can result in a physical description of reality that is mathematically indistinguishable from a mere change in coordinates (a passive diffeomorphism). Therefore, if it is impossible to distinguish between an active diffeomorphism and a passive diffeomorphism, and if the principle of general covariance is taken to be true (that coordinates are indeed unphysical), then it follows that active diffeomorphisms must also be unphysical. Consequently, the spacetime manifold itself must be seen as being unphysical. It is only with respect to events or physical objects (in other words, observable quantities) that locations in spacetime have any meaning at all.

The interpretation here is that spacetime has no real, physical meaning independent of coincidences. What is meant by this is that points in spacetime are only defined by two or more events or objects coinciding, such as two particles passing through the same point in space simultaneously. With this in mind, the existence of more than one solution to a covariant set of field equations is not seen to be problematic. If one metric tensor solves the field equations and yields spacetime paths for multiple observers that coincide at certain proper times for each observer, then \textit{all} metric tensors consistent with the same initial conditions will yield spacetime paths for these observers that coincide at the same proper times. Einstein remarks on this, \textit{``... the requirement of general covariance takes away from space and time the last remnant of physical objectivity''}~\cite{Rovelli2004}.

The principle of general covariance and its interpretation on the meaning (or rather lack of meaning) of spacetime na\"{i}vely appears to be a problem that will be impossible to overcome in analogue gravity models because such models \textit{do} possess an objective, physical analogue of spacetime; there is a preferred reference frame. In order to investigate if or how general covariance manifests in analogue gravity systems, we will need to consider the proper time elapsed by observers in analogue gravity models.

Barcel\'{o} and Jannes~\cite{Barcelo2008} described the physics of `natural' interferometers in analogue relativity models. Such devices would be constructed out of quasiparticles, themselves made up from the particles of the medium under consideration. Observers within these analogue universes, using these interferometers to perform Michelson--Morley type experiments, would find the exact same results that we found in our universe: that the aether is undetectable.

The reason for this is that the medium itself obeys a relativistic wave equation, and thus the resulting kinematics of the medium are subject to the symmetries of the Lorentz group. For example, it has been shown how electromagnetic-like theories based on models that admit a privileged reference frame can, in the low-energy limit, appear to obey Lorentz invariance to internal observers~\cite{1367-2630-16-12-123028}. Consequently, any object constructed from the medium of the analogue universe will inherit the symmetries of the medium, \textit{i.e.} the Lorentz group. Interferometers built this way will possess arms that will shrink in their direction of motion, and thus without any way to measure velocity relative to their aether, observers who only have access to these devices will come to believe the postulates of relativity via Ockham's razor. 

Additional discussions on the emergence of Lorentz symmetry and relativity in physical models that admit a rest frame can be found in the literature. For example, see Liberati, Sonego, and Visser~\cite{Liberati2002}, Volovik~\cite{Volovik2009}, and Nandi~\cite{nandi_2010}.

While quasiparticle interferometers are \textit{sufficient} in demonstrating the emergent relativity of these systems, we would like to ask, are such constructs \textit{necessary} to demonstrate this relativity? An observer within such a system who is free to perform \textit{any} experiment they would like using an interferometer made from a material other than the one used to construct the medium of their universe will be able to infer motion relative to their aether and will not come to believe the postulates of relativity. However, if certain restrictions are placed on which experiments such observers are allowed to perform with non-quasiparticle interferometers, can such observers come to believe the postulates of relativity through their observations?

Operationally speaking, constructing (or even describing how to construct) quasiparticles that would then in turn be used to build devices such as interferometers seems difficult, though such devices would most likely prove to be invaluable additions to the tool-kit of experimental physicists seeking to test analogue relativity systems. To this end, we turn our attention towards devices that are inserted into analogue relativity systems from the laboratory in order to determine if, with certain constraints, such devices can appropriately act as relativistic observers in such analogue systems. With certain restrictions placed on what type of experiments can be conducted, we show here that \textit{sound clocks}~--~the equivalent of light clocks in systems obeying sonic relativity~--~can be used as appropriate relativistic observers for a medium in which sound obeys a relativistic wave equation.

\section{Approach}
\label{Sec: Approach}

As a preface to what we will discuss here, it should be noted that the work carried out here is essentially equivalent to the work performed by Poincar\'{e} at the beginning of the 20\textsuperscript{th} century in his effort to incorporate his philosophy of relativity into the aether based model prevalent at the time. Here, our aether is that of a condensed matter system (such as a large slab of solid matter or a perfect liquid, either of which must be homogeneous and isotropic), and light is replaced by disturbances within our medium: sound. While Poincar\'{e} sought a description for the way in which mechanical bodies naturally behaved in order to obscure detection of the aether, we are, in some sense, attempting to do the reverse: how can we \textit{manufacture} obfuscation of an aether from observers within it in such a way that the observers come to believe the postulates of relativity? What are the minimum constraints that we are required to put on experimental equipment in an aether-based system such that observers who only have access to that equipment have the existence of that aether hidden from them? We ask this questions in the context of analogue gravity models from which we aim to, if possible, make analogies to phenomena in our own universe. However, let it be made abundantly clear that we are not attempting to revive aether-based models as a basis for how our own universe works.

We go through a detailed analysis of the experience of observers who are both stationary and moving with a constant velocity with respect to the medium that they are confined to (which is at rest within the laboratory). The results obtained for stationary observers are in no way surprising, and calculations done from the laboratory frame without considering the technicalities of stationary observers' experience will yield the same results. However, going through the more painstaking analysis is instructive in how to approach the problem of analysing the experience of moving observers. The results found for the experience of moving observers \textit{are} indeed surprising at first glance and are not what is expected from the na\"{i}ve, simple calculation done from the laboratory (which just happens to work for the stationary observers because they share the reference frame of the laboratory, even if they are unaware of it). We will discuss the specifics of the analysis and the restrictions that are necessary as they become important.

\section{Simple sound clocks}
\label{Sec: Simple sound clocks}

Here we describe devices called \textit{sound clocks} for systems in which sound obeys a relativistic wave equation, \textit{i.e.} systems that exhibit sonic relativity. A sound clock is a device that is analogous to the light clock used in thought experiments to show time dilation in special relativity. Sound clocks consist of a clock mechanism out from which an arm, called the \textit{timing arm}, is extended. At both ends of the timing arm, sound can be detected and emitted. To record time, a sound pulse is emitted from the end of the timing arm in contact with the clock mechanism, and at some later point in time, part of the wavefront will intersect the far end of the timing arm and be detected. Immediately following detection, a sound pulse is then emitted back in the same fashion. Upon reception of this second sound pulse~--~what we will herein refer to as the `echo'~--~the clock mechanism advances its reading forward by one tick.

\begin{figure}[tb]
	\centering
	\includegraphics[width=\linewidth]{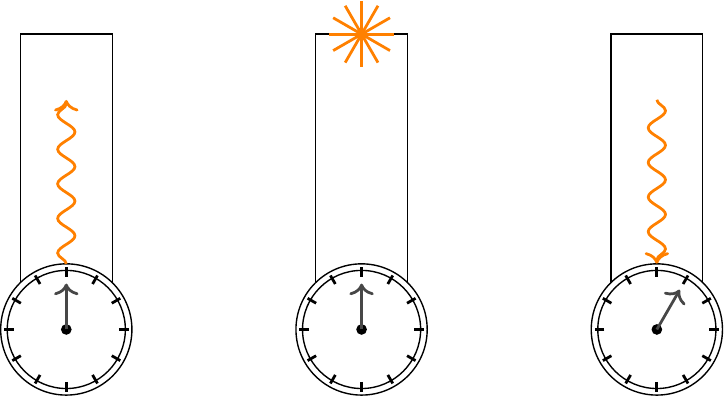}
	\caption[The operation of a sound clock is, in principle, straightforward: a sound pulse is emitted by the clock when an observer wishes to begin recording time, and after travelling to the end of the sound clock arm it is detected, whereupon a new sound pulse (the ‘echo’) is then emitted back towards the central clock mechanism. Upon reception of the echo, the clock reading is advanced forward by one tick. Note that this figure depicts one sound clock at three distinct moments in time as opposed to three sound clocks at one moment in time.]{The operation of a sound clock is, in principle, straightforward: a sound pulse is emitted by the clock when an observer wishes to begin recording time, and after travelling to the end of the sound clock arm it is detected, whereupon a new sound pulse (the ‘echo’) is then emitted back towards the central clock mechanism. Upon reception of the echo, the clock reading is advanced forward by one tick. Note that this figure depicts one sound clock at three distinct moments in time as opposed to three sound clocks at one moment in time.}
	\label{img:Simple sound clock}
\end{figure}

A diagrammatic outline of this process can be seen in \RefFig{img:Simple sound clock}. We could also imagine additional shorter timing arms being present to increment the sound clock by fractions of a tick, but for the sake of clarity in figures, we shall not include these. Observers thus have some sense of \textit{local time} by counting ticks of their clock. However, if an observer wishes to know when an event occurred at some other location, then they must have access to the reading on the clock positioned at that location (for example, by sending a message to the observer at that location and requesting the reading for when a specified event happened).

Let us first consider a single sound clock as observed from the laboratory that is limited in the possible trajectories it can take: it is only allowed to travel in the direction perpendicular to its timing arm. If in the laboratory the length of the sound clock's timing arm is known, and the velocity of the sound clock with respect to the medium is known, then the distance that $n$ sound pulses have travelled can be determined via simple geometric arguments.
\begin{figure}[tb]
	\centering
	\includegraphics[width=\linewidth]{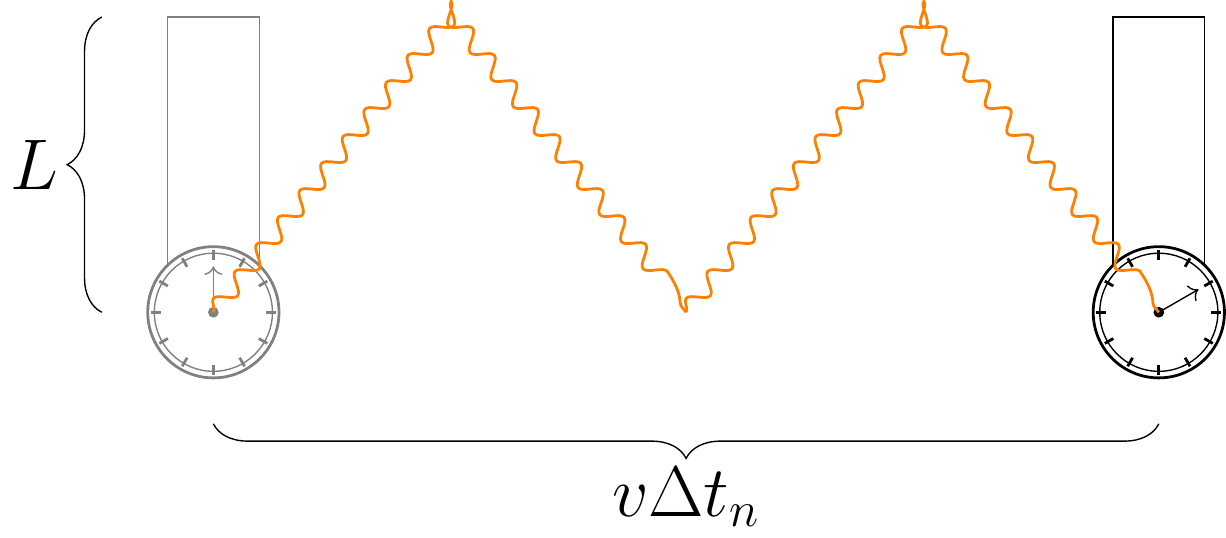}
	\caption[A sound clock travelling with velocity $v$ for some time $\Delta{t_n}$ in which time $n$ sound pulses are emitted and returned to the clock.]{A sound clock travelling with velocity $v$ for some time $\Delta{t_n}$ in which time $n$ sound pulses are emitted and returned to the clock.}
	\label{img:n sound pulse bounces}
\end{figure}

Furthermore, if one knows the speed of sound in the medium, then the time it took for these $n$ sound pulses to trace out their paths can be determined. From \RefFig{img:n sound pulse bounces}, it can be determined that, in the laboratory frame, the total distance travelled by $n$ sound pulses that are emitted and received by a sound clock travelling at velocity $v$ for time $\Delta{t_n}$ ($\Delta{t_n}$ for the duration of time that is required for $n$ sound pulses to be emitted and received) in the direction perpendicular to its timing arm is given by
\begin{equation}
s=\sqrt{\left(2Ln\right)^2+\left(v\Delta{t_n}\right)^2}.
\end{equation}
Dividing the distance that the $n$ sound pulses have travelled by their speed, the speed of sound, $\SoS$, yields the time, $\Delta{t_s}$ ($\Delta{t_s}$ for the time it took to trace out the path of length $s$), that it has taken for these $n$ sound pulses to trace out their paths through space,
\begin{equation}
\Delta{t_s}=\dfrac{s}{\SoS}=\sqrt{\left(\dfrac{2L}{\SoS}n\right)^2+\left(\dfrac{v}{\SoS}\Delta{t_n}\right)^2}.
\end{equation}
Note that $\Delta{t_s}$ and $\Delta{t_n}$ are equal, which is easiest to see for integer values of $n$: when $n$ is integer, a sound pulse has just been detected; in order to detect a sound pulse, the sound clock and the sound pulse that is being detected must be at the same place at the same time. Defining some new variable $\Delta{t}\coloneqq\Delta{t_n}=\Delta{t_s}$ (how long the clock has been recording ticks for), and also defining $\beta\coloneqq{v/\SoS}$, we can determine exactly how long it takes for a sound clock travelling at any velocity less than $\SoS$ to record $n$ ticks of the clock,
\begin{equation}
\Delta{t}=\dfrac{2L}{\SoS}\dfrac{n}{\sqrt{1-\beta^2}}.
\label{Eq:n ticks of any clock}
\end{equation}
The tick frequency, or the period, of any sound clock is therefore given by,
\begin{equation}
T=\dfrac{\Delta{t}}{n}=\dfrac{2L}{\SoS}\dfrac{1}{\sqrt{1-\beta^2}}.
\label{Eq:Sound clock tick frequency}
\end{equation}
The Lorentz factor, $\gamma~=~\sqrt{1-\beta^2}^{-1}$, has appeared for the tick frequency of a moving sound clock. 

Note that for a sound clock at rest with respect to the medium in the laboratory frame we have $v=0$, and thus $\beta=0$, from which we obtain
\begin{equation}
\Delta{t}=\dfrac{2L}{\SoS}n,
\end{equation}
which corresponds to a period of
\begin{equation}
T=\dfrac{2L}{\SoS}.
\label{Eq:Period of stationary sound clock}
\end{equation}
This is exactly the time we expect it to take for $n$ sound pulses to bounce along and back an arm of length $L$ at rest.

\section{Sound clock chains}
\label{Sec: Sound clock chains}
\begin{figure}[tb]
	\centering
	\includegraphics[width=\linewidth]{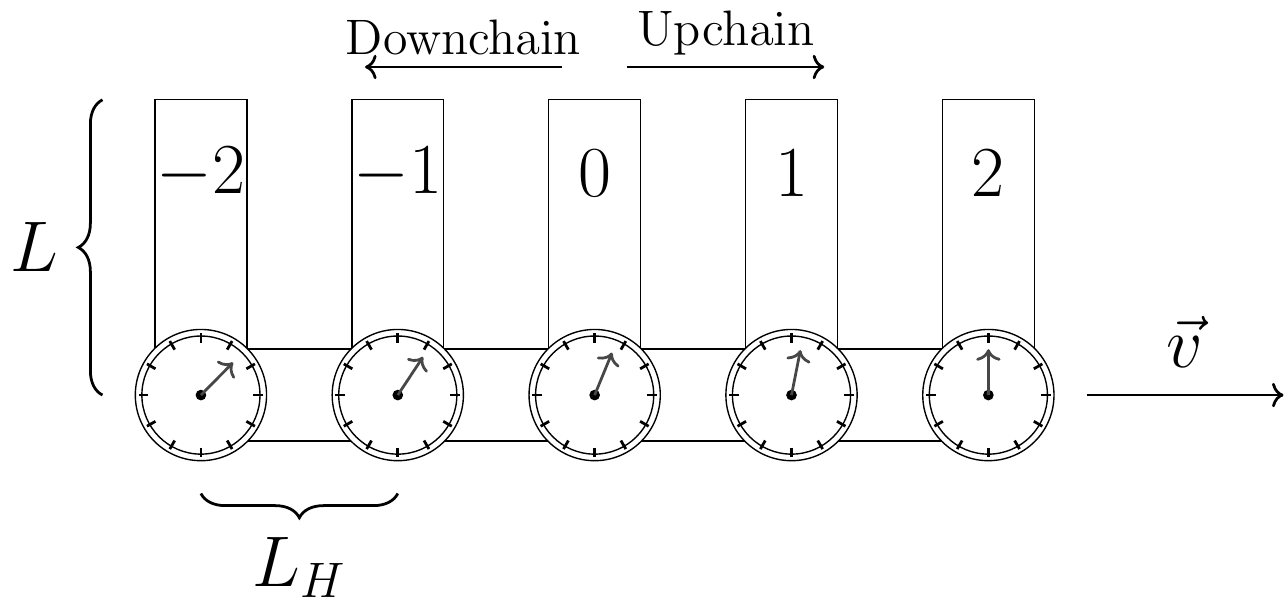}
	\caption[A chain of sound clocks whose clocks have been separated and synchronised under the assumption that they are stationary. The length of the vertical timing arms ($L$) of all sound clocks in the chain is equal, as is the length of the horizontal spacing arms ($L_H$), though the lengths $L_H$ and $L$ are not necessarily equal (only at rest is this the case).]{A chain of sound clocks whose clocks have been separated and synchronised under the assumption that they are stationary. The length of the vertical timing arms ($L$) of all sound clocks in the chain is equal, as is the length of the horizontal spacing arms ($L_H$), though the lengths $L_H$ and $L$ are not necessarily equal (only at rest is this the case).}
	\label{img:Chain of sound clocks}
\end{figure}
Consider now multiple sound clocks at different locations in space. A chain of regularly spaced sound clocks is the easiest such example of multiple sound clocks to consider. The sound clocks that form a chain are connected by arms of tunable length to their neighbours and are synchronised with the use of a sound pulse from some agreed-upon clock (call it the origin clock).

Consider the sound clocks to be labelled with integer values corresponding to how many steps away from the origin clock they are, with the origin clock itself being labelled clock $0$. In general, there can be sound clocks to either side of the origin clock, with clocks on one side possessing positive integer labels, and clocks on the other side possessing negative integer labels. We shall adopt the following convention: when labelled from the laboratory, the sound clock with the largest positive-integer label is at the front of the chain if the chain is in motion; if the chain is stationary, we can freely label either side positive or negative.

Relative to a given sound clock, we call clocks closer to the front of a moving chain `upchain' and clocks closer to the back of a moving chain `downchain': from the definition of the labelling scheme outlined above this means that the direction `upchain' is parallel to the sound clock chain's velocity vector, and the direction `downchain' is anti-parallel to the sound clock chain's velocity vector. The labelling convention is shown in \RefFig{img:Chain of sound clocks}.

We wish for all of the sound clocks within a given chain to share a common coordinate system, so sound clocks within a chain must tick at the same frequency. The timing arms of adjacent sound clocks are assumed to be exactly parallel, and so to fulfil the requirement that they tick synchronously, the timing arms must be of equal length. This requirement and others will be discussed further in what follows.

\subsection{Calibrating clock separation}
\label{Subsec: Calibrating clock separation}

A chain of sound clocks can be seen in \RefFig{img:Chain of sound clocks}, where $L$ refers to the length of the `vertical' arms (\textit{i.e.} the timing arms), and $L_H$ is the length of the horizontal arm separating sound clocks, which we also call the \textit{spacing arms}. Note that for what follows the terms `vertical arms' and `horizontal arms' are interchangeable with `timing arms' and `spacing arms', respectively. The vertical arm of each sound clock is used to measure time directly in the same method as described for the single sound clock in \RefSec{Sec: Simple sound clocks}, while the horizontal arm is used to space the sound clocks in an appropriate manner. For the purposes of our discussions here, the timing arms and spacing arms are considered to always be perpendicular to one another. We focus on the case where chains of sound clocks are \textit{only} allowed to travel in the direction of the axis in which they are connected.

Observers who possess sound clocks and who are limited in their ability to make measurements as previously detailed have no ability to detect motion with respect to their medium. Any inertial motion within an analogue relativity system should therefore be indistinguishable from rest since observers who only possess sound clocks have no way to tell the difference between zero velocity with respect to the medium and constant non-zero velocity with respect to the medium.

Observers travelling with constant velocity separate sound clocks within a chain by simultaneously sending sound pulses along their timing arms and spacing arms. Believing themselves to be at rest, when the spacing arms are tuned to such a length that the two sound pulses return simultaneously, observers believe that the separation of their sound clocks (\textit{i.e.} the spacing-arm length) is exactly the same as the length, $L$, of their timing arms. This belief happens to be true when a chain of sound clocks is actually at rest, but this is not the case when a chain of sound clocks is moving with a constant velocity.

\begin{figure}[tb]
	\centering
	\includegraphics[width=\linewidth]{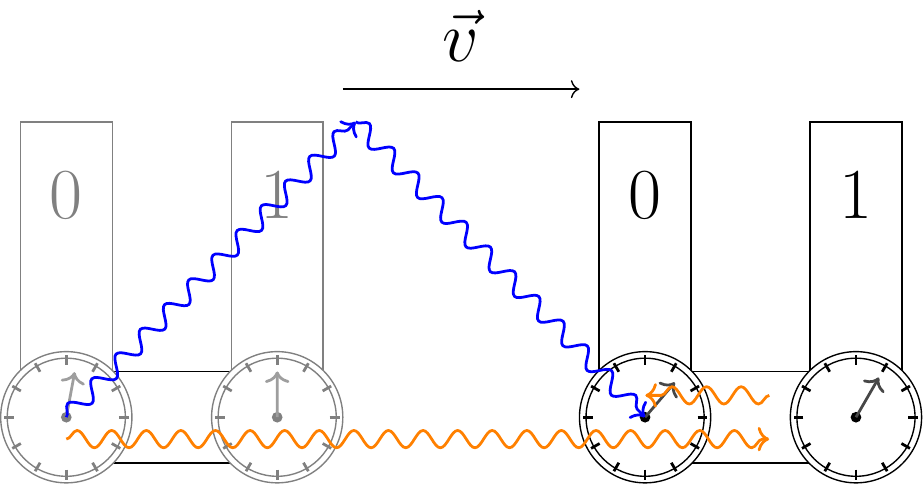}
	\caption[The spacing arms between adjacent sound clocks in moving sound clock chains must be shorter than the timing arms by a factor of $\gamma$, to ensure that sound pulses that are emitted simultaneously along both arms also return simultaneously. Note that the readings of both clocks are advanced by exactly one tick when the sound pulses return (this is obscured partially for clock $0$).]{The spacing arms between adjacent sound clocks in moving sound clock chains must be shorter than the timing arms by a factor of $\gamma$, to ensure that sound pulses that are emitted simultaneously along both arms also return simultaneously. Note that the readings of both clocks are advanced by exactly one tick when the sound pulses return (this is obscured partially for clock $0$).}
	\label{img:Sound clocks length contraction as seen by laboratory}
\end{figure}

By what distance should sound clocks travelling at a constant velocity be separated in order for this simultaneity condition to hold? Consider a simple chain of two sound clocks as seen in \RefFig{img:Sound clocks length contraction as seen by laboratory}. The observer located at one of the sound clocks is able to adjust the separation distance of the two sound clocks by extending the arm that connects them or by reeling it back in.

Within a chain of sound clocks, the path taken by the sound pulse within the timing arms on its outbound journey and its inbound journey is symmetric for any motion perpendicular to the timing arms. This, however, is not the case for a sound pulse in the spacing arms. For sound pulses propagating between clocks in the spacing arms, there are two distinct paths taken when the clocks are travelling with non-zero velocity. As can be seen in \RefFig{img:Sound clocks length contraction as seen by laboratory}, there is a downchain journey for which the sound pulse is travelling in the opposite direction to the sound clocks, and there is an upchain journey for which the sound pulse is travelling in the same direction as the sound clocks. The downchain journey, as seen in \RefFig{img:Sound clocks length contraction as seen by laboratory}, takes less time to complete as $v$ grows, while the upchain journey takes longer. The time it takes for a sound pulse to travel between two adjacent clocks downchain ($\Delta{t_d}$, which is greater than zero) and the time it takes to travel between two adjacent clocks upchain ($\Delta{t_u}$, which is greater than zero), as shown in \RefFig{img:Sound clocks length contraction as seen by laboratory}, can be seen to obey the following relationships for a separation length of $L_H$,
\begin{align}
L_{H}-v\Delta{t_d}&=\SoS\Delta{t_d},\\
\therefore~\Delta{t_d}&=\dfrac{L_H}{\SoS+v},
\label{Eq:Downchain time}
\end{align}
and,
\begin{align}
L_{H}+v\Delta{t_u}&=\SoS\Delta{t_u},\\
\therefore~\Delta{t_u}&=\dfrac{L_H}{\SoS-v}.
\label{Eq:Upchain time}
\end{align}
For simultaneously emitted timing and synchronisation sound pulses to be detected simultaneously, we don't need to know exactly how long it takes the sound pulse to make either the downchain or upchain journey alone. We know that, by construction, the sum of the upchain and downchain times must be the same as the time for which one tick of the clock occurs as dictated by the timing arm, and one tick of the clock is given by \RefEq{Eq:n ticks of any clock} with $n=1$. With the length of the vertical arm labelled $L$ we have the relationship
\begin{equation}
\Delta{t}=\Delta{t_d}+\Delta{t_u}=\dfrac{2L}{\SoS}\dfrac{1}{\sqrt{1-\beta^2}}.
\end{equation}
Substituting in the relationships for $\Delta{t_d}$ and $\Delta{t_u}$ allows us to solve for the separation distance, $L_H$, which we find to be
\begin{equation}
L_H=\dfrac{L}{\gamma}.
\label{Eq:Length contraction in moving sound clock chains}
\end{equation}

\subsection{Synchronisation of clocks}
\label{Subsec: Synchronisation of clocks}

When the origin clock (clock $0$) in a sound clock chain begins to record time, it simultaneously sends sound pulses along every arm connected to it. The sound pulse sent down its own timing arm is used to advance its own clock, whereas the pulses sent along the sound clock chain via the spacing arms are used to synchronise directly adjacent clocks (these are the synchronisation pulses). Upon receiving the synchronisation pulse, a given clock will simultaneously begin to record its own time and propagate the synchronisation pulse further along the chain. By this manner, all of the clocks in the chain can be synchronised with respect to the origin clock.

We can construct expressions for the time taken for the synchronisation pulse emitted from the origin clock to travel to another clock: $\Delta{t^{+}_{s}}=\left|k\right|\Delta{t_u}$ is the time it takes for the synchronisation pulse to reach clock $k$ ($k$ steps in the upchain direction), while $\Delta{t^{-}_{s}}=\left|k\right|\Delta{t_d}$ is the time it takes for the synchronisation pulse to to reach clock $-k$ ($k$ steps in the downchain direction). When at rest, $\Delta{t^{+}_{s}}$ and $\Delta{t^{-}_{s}}$ are equal. The relationships for the downchain and upchain synchronisation times can be combined into a single expression, and substituting in the expression for $L_H$ from \RefEq{Eq:Length contraction in moving sound clock chains} we find
\begin{equation}
\Delta{t^{\pm}_s}=\dfrac{L}{\SoS}\sqrt{\dfrac{1\pm\beta}{1\mp\beta}}\left|{k}\right|.
\label{Eq:Synchronisation time for clock i}
\end{equation}
Despite the fact that the sound clocks are travelling in a medium with a preferred reference frame, the relativistic Doppler factor,
\begin{equation}
D = \sqrt{\dfrac{1+\beta}{1-\beta}},
\end{equation}
has appeared instead of the non-relativistic one, where $\beta$ is the fractional speed of the sound clock chain with respect to sound. This is a result of observers within the sound clock chain setting the separation between adjacent clocks in such a way that the simultaneity of returned sound pulses occurs. In setting their separation in such a way, they have not only made their chain appear to exhibit the relativistic phenomenon of length contraction to observers in the laboratory, but they have also made their system appear to display the relativistic Doppler shift in regards to how long it takes sound to propagate between adjacent clocks to observers within the laboratory.

We now know how long it takes for any given clock, $k$, to tick $n$ times ($n_k$) once it begins its clock: this is given by \RefEq{Eq:n ticks of any clock}. We also know how long it takes for clock $k$ to start its clock with respect to some initial clock (\textit{i.e.} the time until it receives the synchronisation pulse): this is given by the appropriate choice of synchronisation time from \RefEq{Eq:Synchronisation time for clock i}. For $n$ ticks of the $k$\textsuperscript{th} clock ($n_k$), the total amount of time that has transpired since the lead clock first began recording time is simply given by the sum of these two times.

For a clock $a$ in the upchain direction we have the expression
\begin{equation}
t=\dfrac{L}{\SoS}\sqrt{\dfrac{1+\beta}{1-\beta}}\left|{a}\right|+\dfrac{2L}{\SoS}\dfrac{n_a}{\sqrt{1-\beta^2}},
\label{Eq:Total time elapsed upchain}
\end{equation}
and for a clock $b$ in the downchain direction we have the expression
\begin{equation}
t = \dfrac{L}{\SoS}\sqrt{\dfrac{1-\beta}{1+\beta}}\left|{b}\right| + \dfrac{2L}{\SoS}\dfrac{n_b}{\sqrt{1-\beta^2}}.
\label{Eq:Total time elapsed downchain}
\end{equation}
Let us also rearrange these equations for the number of ticks recorded by any clock post synchronisation. Upchain, the number of ticks recorded by clock $a$ after time $t$ is
\begin{equation}
n_a=\dfrac{{\SoS}t\sqrt{1-\beta^2}}{2L}-\dfrac{\left|{a}\right|}{2}(1+\beta),
\label{Eq:Returned sound pulses for any clock upchain}
\end{equation}
while downchain the number of ticks recorded by clock $b$ after time $t$ is
\begin{equation}
n_b=\dfrac{{\SoS}t\sqrt{1-\beta^2}}{2L}-\dfrac{\left|{b}\right|}{2}(1-\beta).
\label{Eq:Returned sound pulses for any clock downchain}
\end{equation}
Adding to \RefEq{Eq:Returned sound pulses for any clock upchain} and \RefEq{Eq:Returned sound pulses for any clock downchain} the offsets that the observers located at clocks $a$ and $b$ will add to their clocks to account for the believed synchronisation time ($\left|a\right|/2$ and $\left|b\right|/2$, respectively) yields the proper clock reading, denoted $\nu$, for clocks that are upchain and downchain, respectively. Note that, by the labelling scheme, we can express the proper clock reading of \textit{any} clock in the chain, $k$ (where $k$ is any integer), at an instant in time with a single formula,
\begin{equation}
\nu_k\coloneqq{n_k}+\dfrac{\left|k\right|}{2}=\dfrac{{\SoS}t\sqrt{1-\beta^2}}{2L}-\dfrac{k}{2}\beta.
\label{Eq:Proper clock reading}
\end{equation}
From this equation we can rearrange for $t$ again, yielding
\begin{equation}
t=\dfrac{L}{\SoS}\left(\left|k\right|+k\beta\right)\gamma+\dfrac{2L}{\SoS}n_k\gamma=\dfrac{2L}{\SoS}\left(\nu_k+\dfrac{k}{2}\beta\right)\gamma.
\label{Eq:Total time elapsed}
\end{equation}
\RefEq{Eq:Total time elapsed upchain} and \RefEq{Eq:Total time elapsed downchain} are limiting cases of \RefEq{Eq:Total time elapsed} and are now superfluous.

From \RefEq{Eq:Proper clock reading} or \RefEq{Eq:Total time elapsed}, one can obtain the useful relationships for the difference in time as a function of the difference of proper clock reading of any clock or clocks,
\begin{equation}
\Delta{t}\left(\Delta{\nu}\right)=t(\nu_l)-t(\nu_k) = \dfrac{2L}{\SoS}\gamma\Delta{\nu} + \dfrac{L}{\SoS}\gamma\beta\left(l-k\right),
\label{Eq:Difference in time as a function of difference in proper clock reading}
\end{equation}
and for the difference in proper clock reading for any clock or clocks as a function of a difference in time,
\begin{equation}
\Delta{\nu}\left(\Delta{t}\right)=\nu_l(t_1)-\nu_k(t_0)=\dfrac{\SoS}{2L\gamma}\Delta{t} + \dfrac{1}{2}\beta\left(k-l\right).
\label{Eq:Difference in proper clock reading as a function of difference in time}
\end{equation}
These two relationships, \RefEq{Eq:Difference in proper clock reading as a function of difference in time} and \RefEq{Eq:Difference in time as a function of difference in proper clock reading}, are of crucial importance in \RefSec{Sec:Relativistic effects observed by stationary sound clocks} and \RefSec{Sec:Relativistic effects observed by moving sound clocks}. It is important to note that the quantity $\Delta{\nu}$ present in \RefEq{Eq:Difference in time as a function of difference in proper clock reading} and \RefEq{Eq:Difference in proper clock reading as a function of difference in time} is entirely general and can correspond to the difference in proper clock reading as recorded by a single clock ($k=l$), which must of course occur at two different instances in time ($\Delta{t}\neq{0}$), or to the difference in proper clock reading as recorded by different clocks ($k\neq{l}$), which can be calculated for any difference in time.

When observers~--~moving or not~--~request information from another observer on their chain for a believed simultaneous point in time, they will request information recorded at the same proper clock reading (\textit{i.e.} $\Delta{\nu}=0$). This is because all observers in inertial motion believe themselves to be at rest and so believe that all synchronised clocks possess the same proper clock readings at a given point in time. In reality, a moment in time is given by $\Delta{t}=0$ for which, in general, different clocks will not read a difference in proper clock reading of $\Delta{\nu}=0$ (which only happens to be true for chains of sound clocks at rest).

\begin{figure}[tb]
	\centering
	\includegraphics[width=\linewidth]{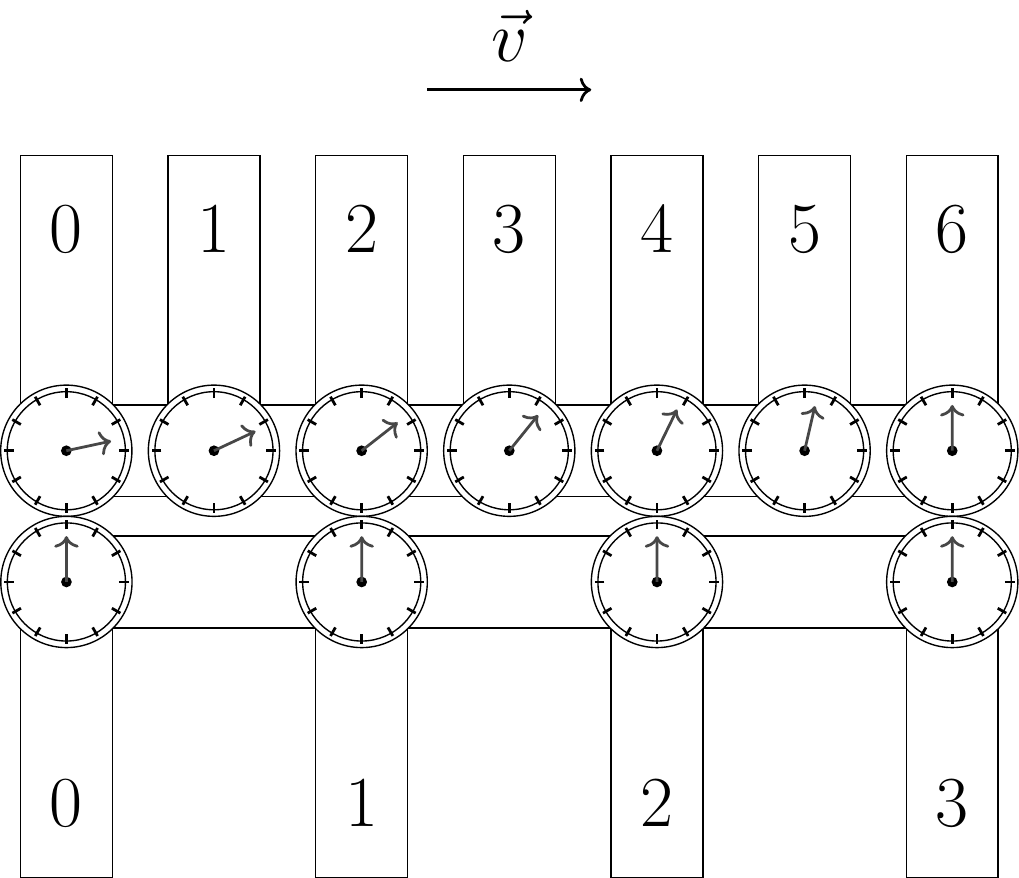}
	\caption[Two chains of sound clocks instantaneously adjacent to one another after having carried out their calibration and synchronisation procedures. The sound clock chain at the \textit{bottom} is stationary, while the sound clock chain at the \textit{top} is travelling with a fractional velocity of $\beta=\sqrt{3}/2$, corresponding to a Lorentz factor of $\gamma=2$. Simultaneity requirements lead to the moving chain possessing asynchronous clocks that are separated by a distance of $L/\gamma$, where $L$ is the length of the vertical timing arms.]{Two chains of sound clocks instantaneously adjacent to one another after having carried out their calibration and synchronisation procedures. The sound clock chain at the \textit{bottom} is stationary, while the sound clock chain at the \textit{top} is travelling with a fractional velocity of $\beta=\sqrt{3}/2$, corresponding to a Lorentz factor of $\gamma=2$. Simultaneity requirements lead to the moving chain possessing asynchronous clocks that are separated by a distance of $L/\gamma$, where $L$ is the length of the vertical timing arms.}
	\label{img:Synchronised chains}
\end{figure}

\RefFig{img:Synchronised chains} shows two chains of sound clocks, one at rest and one travelling with a velocity of $v$ parallel to the stationary chain, at an instant in time. It can be seen that, as per \RefEq{Eq:Difference in proper clock reading as a function of difference in time}, the stationary chain ($\beta=0$) possesses clocks that all have the same reading at an instant in time, whereas the moving chain possesses clocks whose readings differ as a function of their separation from the origin clock.

\section{Relativistic effects observed by stationary sound clocks}
\label{Sec:Relativistic effects observed by stationary sound clocks}

Consider the following scenario: a moving chain of sound clocks passes by a stationary chain of sound clocks with a velocity of ${v}$, which corresponds to a fractional speed of ${\beta}$, with respect to sound. The moving chain travels in the direction parallel to its own spacing arms, and passes by the stationary chain parallel to it and close enough to it that the travel time for sound between clock faces on adjacent chains is small with respect to the time it takes clocks on either chain to tick. Observers located at each clock can only record the information accessible from their immediate surroundings: they can read their own clock face, they can count how many clock faces they have passed by on the adjacent chain, and they can read the value recorded on clocks in the adjacent chain when they are sufficiently close (\textit{i.e.} next to them). Observers within a chain then have to talk to one another and exchange their own measurements in order to come to some understanding of whatever experiment they conducted.

As we shall demonstrate, observers located on the stationary chain of sound clocks determine that the sound clocks within the moving chain appear to be both separated by a shorter distance and tick less frequently than their own sound clocks. This is in keeping with what is seen in the laboratory. Later, when we consider measurements made by observers on a moving sound clock chain of a stationary chain we find that, contrary to what occurs in the laboratory, moving observers also determine that the clocks in the stationary chain are separated by a shorter distance than their own and ticks less frequently than their own. This only happens when both chains are treated equally in that observers are only allowed to use their own clocks as time references, observers can only signal with sound pulses, and observers have no means by which to detect motion with respect to the medium that they are embedded within.

\subsection{Time dilation as seen by stationary observers}
\label{Subsec:Time dilation as seen by stationary observers}

\begin{figure}[tb]
	\centering
	\includegraphics[width=\linewidth]{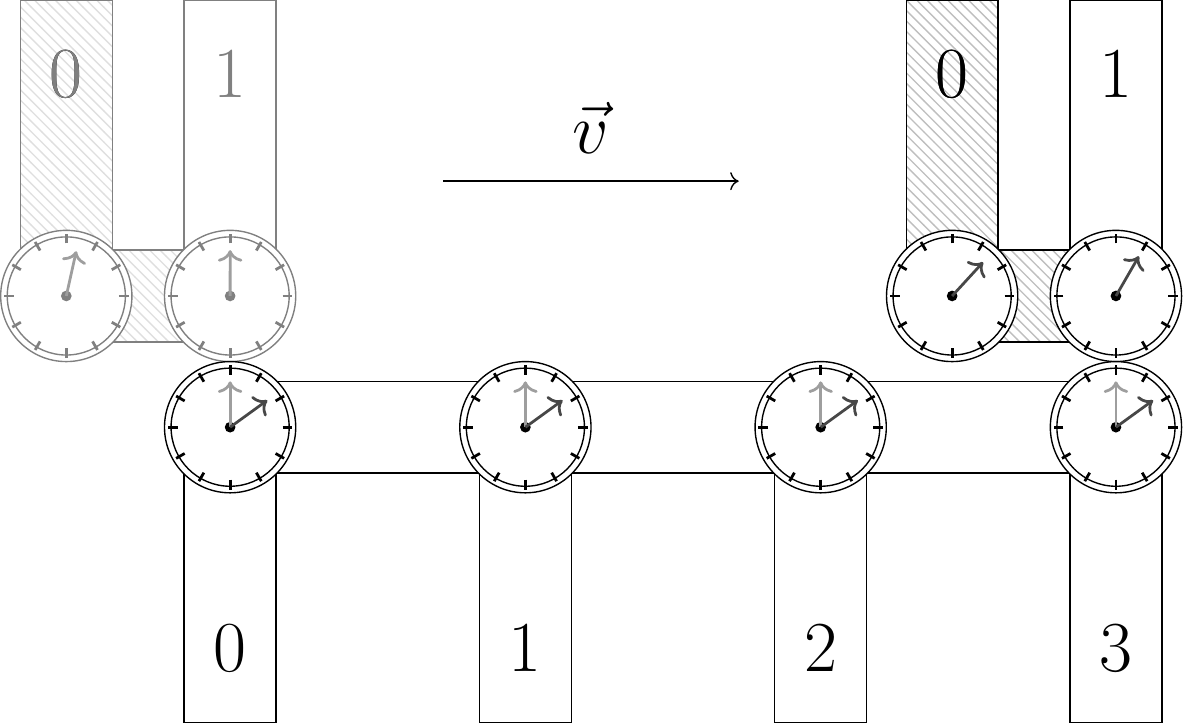}
	\caption[The observers in the stationary chain focus only on the lead sound clock of the moving chain. The moving chain has a fractional velocity of $\beta=3/\sqrt{13}$, with respect to sound, corresponding to a Lorentz factor of $\gamma=\sqrt{13}/2\approx1.8$. With this velocity, the moving chain of sound clocks travels a distance of $3L$ in the laboratory for every tick of its clocks. In this example, the clocks labelled $k$ and $l$ in the stationary chain are clocks $0$ and $3$, respectively.]{The observers in the stationary chain focus only on the lead sound clock of the moving chain. The moving chain has a fractional velocity of $\beta=3/\sqrt{13}$, with respect to sound, corresponding to a Lorentz factor of $\gamma=\sqrt{13}/2\approx1.8$. With this velocity, the moving chain of sound clocks travels a distance of $3L$ in the laboratory for every tick of its clocks. In this example, the clocks labelled $k$ and $l$ in the stationary chain are clocks $0$ and $3$, respectively.}
	\label{img:Sonic time dilation chain B}
\end{figure}

Imagine that observers within the stationary chain of sound clocks decide to focus on the lead sound clock of the moving chain as it passes by, as seen in \RefFig{img:Sonic time dilation chain B}. The first experiment they wish to conduct is:

\textit{``How many times do moving clocks appear to tick for every tick of stationary ones?''}

To determine how many times sound clocks in a measuring chain believe themselves to tick for every one tick of a sound clock in a different chain, we use the following procedure:
\begin{enumerate}
	\item Determine the separation in time, $\Delta{t}$, for some clock, $z$, in the chain that is being measured to increment its clock reading once. This is obtained using \RefEq{Eq:Difference in time as a function of difference in proper clock reading} with $z=l=k$ and $\Delta{\nu_z}=1$.\label{Step:s1}
	\item Determine which clocks, $k$ and $l$, in the chain performing measurements clock $z$ is next to at two points in time separated by $\Delta{t}$. This is done by determining how far clock $z$ has moved in $\Delta{t}$ as calculated in \RefStep{Step:s1}.\label{Step:s2}
	\item Determine the proper clock reading on clock $k$ when the clock that is being measured, $z$, is next to it, and determine the proper clock reading on clock $l$ when the clock that is being measured, $z$, is next to it. The difference in these two proper clock times is how many ticks observers within a measuring chain believe to have occurred for them for one tick of the clock they were measuring. In other words, evaluate \RefEq{Eq:Difference in proper clock reading as a function of difference in time} for clocks $k$ and $l$ as determined in \RefStep{Step:s2} using time difference obtained in \RefStep{Step:s1}. The difference in these proper clock readings gives the perceived number of ticks that have transpired in the chain performing measurements for one tick of the clock being measured.
\end{enumerate}
Consider that some clock, $k$, in the stationary chain is next to a clock in the moving chain, $z$, when the moving clock has a proper clock reading $\nu_z\moch$ (where the superscript `M' denotes that the quantity pertains to the moving chain). At some later point in time, clock $z$ in the moving chain has moved next to some other clock in the stationary chain, $l$, at the moment that clock $z$ advances its proper clock reading forward one tick to $\nu_z\moch+1$. From \RefEq{Eq:Difference in time as a function of difference in proper clock reading}, the time it takes for a given clock in the moving chain, $z$, to tick once ($\Delta{\nu_z\moch}=1$) is $2L{\gamma}/\SoS$ where ${\gamma}$ is the sonically relativistic Lorentz factor of the moving clock. The distance covered by a moving clock ticking once is then given by
\begin{equation}
x\moch = {v}{\Delta{t}} = {v}\dfrac{2L}{\SoS}{\gamma} = 2L\gamma\beta.
\end{equation}
Stationary clocks are separated by a length of $L$, so the number of stationary-clock spacing-arm lengths that the moving clock has travelled in this time is given by $bL$, where $b$ is just some number. Equating these two distances and solving for $b$, we find,
\begin{equation}
b = 2\gamma\beta.
\label{Eq:Stationary experiment Number of clocks in stationary chain that the moving chain has moved over in one moving tick}
\end{equation}
We have then that $l=k+b$. What is the difference in the proper clock reading, $\Delta{\nu\stch}~=~\nu_l\stch~-~\nu_k\stch$ (the superscript `S' denotes that the quantity pertains to the stationary chain), corresponding to the difference in time, $\Delta{t}$, that it takes for clock $z$ in the moving chain to travel between clocks $k$ and $l$ in the stationary chain? From \RefEq{Eq:Difference in proper clock reading as a function of difference in time} (with $\beta=0$ as we are considering the proper clock reading difference of the stationary chain), we find the difference in proper clock readings to be,
\begin{align}
\Delta{\nu\stch}(\Delta{t}) = {\nu_l\stch}(t_1) - {\nu_k\stch}(t_0) = {\gamma}.
\label{Eq:Stationary observers measure time dilation}
\end{align}
Observers in the stationary chain determine that their clocks have all ticked ${\gamma}$ times for one tick of the moving clock. This is, in fact, true. In the laboratory, all clocks within the stationary chain possess the same proper clock reading at the same instant in time, and all stationary clocks have indeed ticked ${\gamma}$ times for one tick of the moving clock.

\subsection{Length contraction as seen by stationary observers}
\label{Subsec:Length contraction as seen by stationary observers}

Now imagine that observers located in the stationary chain decide to focus on more than just the first sound clock of the moving chain. The next experiment they wish to perform is:

\textit{``How are clocks in the moving chain spaced, relative to clocks in the stationary chain?’’}
\begin{figure}[tb]
	\centering
	\includegraphics[width=\linewidth]{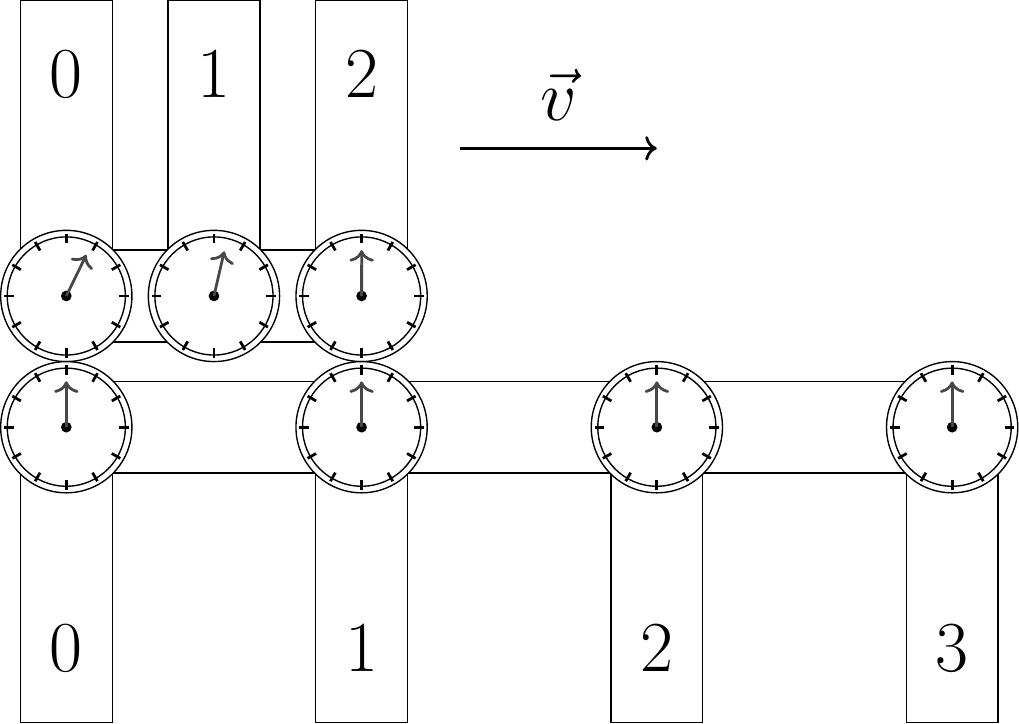}
	\caption[The observers in the stationary chain of sound clocks determine the separation of sound clocks in the moving chain by determining which of their own clocks are simultaneously next to a given pair of clocks in the moving chain. From this information, they are able to determine how many spacing arms in the stationary chain are simultaneously parallel to a given number of spacing arms in their own chain, and with the knowledge that their own spacing arms are of length L, they can determine how long the spacing arms in the moving chain are. The moving chain is travelling with a fractional velocity of $\beta=\sqrt{3}/2$, corresponding to a Lorentz factor of $\gamma = 2$. In this example, clocks $k$ and $l$ in the stationary chain are clocks $0$ and $1$, respectively.]{The observers in the stationary chain of sound clocks determine the separation of sound clocks in the moving chain by determining which of their own clocks are simultaneously next to a given pair of clocks in the moving chain. From this information, they are able to determine how many spacing arms in the stationary chain are simultaneously parallel to a given number of spacing arms in their own chain, and with the knowledge that their own spacing arms are of length L, they can determine how long the spacing arms in the moving chain are. The moving chain is travelling with a fractional velocity of $\beta=\sqrt{3}/2$, corresponding to a Lorentz factor of $\gamma = 2$. In this example, clocks $k$ and $l$ in the stationary chain are clocks $0$ and $1$, respectively.}
	\label{img:Sonic length contraction chain B}
\end{figure}

When we want to determine the length of an object in a laboratory (at least for any experiment that takes place over reasonable distances and times) we will typically find the positions of both ends of the object in question at an instant in time and then determine the separation of these points in space. This is done with respect to some fixed coordinate system parallel to the object (\textit{e.g.} a ruler). The definition of `an instant in time', a notion that is crucial in operationally determining lengths, is a velocity-dependant quantity for observers who only have access to sound clocks. The only measure of time that observers with sound clocks have access to is their proper clock reading as given by \RefEq{Eq:Proper clock reading}. Within a given chain, observers at different clocks will possess a different proper clock reading at a given point in time if their chain is travelling with respect to the medium (\textit{i.e.} $\beta\neq0$ in \RefEq{Eq:Difference in proper clock reading as a function of difference in time}). To determine how observers with sound clocks measure distance, we must ask what measurements they make when \textit{their} clocks have the same readings. The procedure we follow to determine what distance one chain of sound clocks measures another chain's spacing arms to be is:
\begin{enumerate}
	\item Determine which two clocks, $k$ and $l$ (where $k \neq l$), in a given chain are going to be used to perform measurements on lengths in another chain.
	\item Determine the difference in time that is required for these two recording clocks to have the same proper clock reading, \textit{i.e.} determine which two times correspond to $\nu_k = \nu_l$ using \RefEq{Eq:Proper clock reading}. The difference in time between these two clock readings is given by \RefEq{Eq:Difference in time as a function of difference in proper clock reading} for $\Delta\nu=0$ and $k\neq{l}$.
	\item Determine which clock, $k^\prime$, in the chain that is being measured is next to clock $k$ in the chain performing measurements at the time value corresponding to $\nu_k$, and which clock, $l^\prime$, in the chain that is being measured is next to clock $l$ in the chain that is performing measurements at the time value corresponding to $\nu_l$.
	\item Determine how many spacing arms, $b$, separate the clocks in the chain performing measurements ($\left|b\right|=\left|l-k\right|$). Determine how many spacing arms, $b^\prime$, separate the clocks in the chain being measured ($\left|b^\prime\right|=\left|l^\prime-k^\prime\right|$). Using this information, determine how many spacing arms in the chain being measured simultaneously appear to be parallel to one spacing arm in the chain performing the measurements.
\end{enumerate}
Consider that, at some time, one of the clocks in a moving chain, $k^\prime$, is next to one of the clocks (call it $k$) in the stationary chain: we shall label this time $t(\nu_k\stch)$. Which clock in the moving chain, $l^\prime$, is next to some other clock in the stationary chain, call it $l=k+b$, when clock $l$ has the same proper clock reading as clock $k$ at time $t(\nu_k\stch)$? 
From \RefEq{Eq:Difference in time as a function of difference in proper clock reading}, with $\beta=0$, the difference in time between the moments when clocks $k$ and $l$ have the same reading is $0$. That is to say,
\begin{equation}
t(\nu_{k}\stch) = t(\nu_{l}\stch).
\end{equation}
As expected, the clocks in the stationary chain possess the same proper clock reading at the same instant in time. This means that the moving chain has not moved relative to the stationary chain when clocks $k$ and $l$ perform their measurements at equal clock readings, as can be seen in \RefFig{img:Sonic length contraction chain B}, for $b=1$.

Clock $l$ is $b$ spacing arms away from clock $k$, and the spacing arms in the stationary chain have length $L$; therefore clocks $k$ and $l$ are separated by a distance of $bL$. Clock $k^\prime$ in the moving chain is next to clock $k$ in the stationary chain, while simultaneously clock $l^\prime$ in the moving chain is next to clock $l$ in the stationary chain; the number of spacing arms separating clocks $k^\prime$ and $l^\prime$ in the moving chain is labelled $b^\prime$. Clocks in the moving chain are separated by spacing arms of length $L/{\gamma}$, so we have the equality
\begin{equation}
bL = b^\prime\dfrac{L}{\gamma}.
\end{equation}
This leads to the relationship
\begin{equation}
b^\prime = {\gamma}b.
\label{Eq:Stationary chain measures number of moving clocks}
\end{equation}
Directly adjacent clocks in the stationary chain (when $b=1$ in \RefEq{Eq:Stationary chain measures number of moving clocks}) will (correctly) conclude that, parallel to the single spacing arm that separates them, there are $\gamma$ spacing arms in the moving chain.

Furthermore, if one were to extend the formalism and imagine that there were clocks at every point in space along a chain of sound clocks, the hypothetical clock with label $k+1/{\gamma}$ would be next to some clock in the moving chain, whose neighbour would be next to clock $k$ in the stationary chain.

\section{Relativistic effects observed by moving sound clocks}
\label{Sec:Relativistic effects observed by moving sound clocks}

Consider that observers located on a moving chain of sound clocks wish to perform the same experiments as the stationary chain did in \RefSec{Sec:Relativistic effects observed by stationary sound clocks}.

To reiterate: a moving chain of sound clocks passes by a stationary chain of sound clocks with a velocity of $v$, which corresponds to a fractional speed of ${\beta}$, with respect to sound. The moving chain travels in the direction parallel to its own spacing arms and passes by the stationary chain parallel to it and close enough to it that the time it takes sound to propagate between the two chains is small with respect to the time it takes clocks on either chain to tick. Observers located at each clock only have at hand the information accessible from their immediate surroundings; they can count how many clock faces they pass by on the adjacent chain, and they can read the value off of adjacent clock faces when they are sufficiently close. Observers within the chain then have to talk to one another and exchange their own measurements in order to come to some understanding of whatever experiment they conducted.

As foreshadowed in \RefSec{Sec:Relativistic effects observed by stationary sound clocks}, we find that observers located on moving chains of sound clocks find that sound clocks within a stationary chain appear to be both separated by a shorter distance and tick less frequently than their own sound clocks. It is not obvious that this result should appear, and in fact it only does so when the observers are constrained to use their own clocks as time references, can only signal with sound pulses, and cannot detect motion with respect to the medium that they are embedded within. These constraints lead observers to ask for measurements made at the wrong laboratory time when they wish to aggregate and compare data recorded `simultaneously' because the clocks in the moving chain do not actually possess the same reading at an instant in time.

\subsection{Time dilation as seen by moving observers}
\label{Subsec:Time dilation as seen by moving observers}

\begin{figure}[tb]
	\centering
	\includegraphics[width=\linewidth]{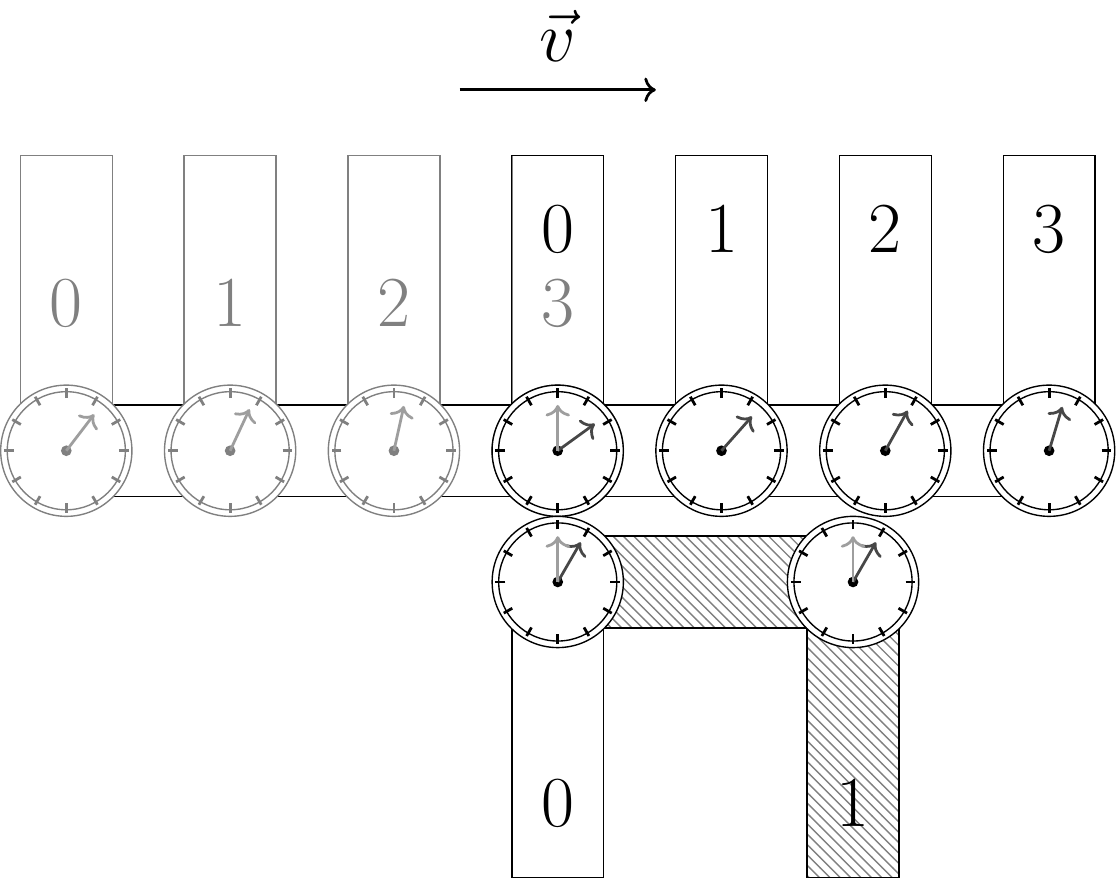}
	\caption[The observers in the moving chain focus only on the lead sound clock of the stationary chain in order to determine how many times their own clocks tick for every one tick of a clock in the moving chain. Due to the asynchronicity of clocks in the moving chain, the observers within the moving chain come to the incorrect conclusion that their clocks tick $\gamma$ times for every $1$ tick of a clock in the stationary chain. The moving chain is travelling with a fractional velocity of $\beta=3/\sqrt{13}$, with respect to sound, corresponding to a Lorentz factor of $\gamma=\sqrt{13}/2\approx1.8$. In this example, the clocks labelled $k$ and $l$ in the moving chain are clocks $3$ and $0$, respectively.]{The observers in the moving chain focus only on the lead sound clock of the stationary chain in order to determine how many times their own clocks tick for every one tick of a clock in the moving chain. Due to the asynchronicity of clocks in the moving chain, the observers within the moving chain come to the incorrect conclusion that their clocks tick $\gamma$ times for every $1$ tick of a clock in the stationary chain. The moving chain is travelling with a fractional velocity of $\beta=3/\sqrt{13}$, with respect to sound, corresponding to a Lorentz factor of $\gamma=\sqrt{13}/2\approx1.8$. In this example, the clocks labelled $k$ and $l$ in the moving chain are clocks $3$ and $0$, respectively.}
	\label{img:Sonic time dilation_chain_A}
\end{figure}

Imagine that observers in the moving chain decide to focus on the lead sound clock of the stationary chain as they pass it, as seen in \RefFig{img:Sonic time dilation_chain_A}. The first experiment they wish to conduct is:

\textit{``How many times do stationary clocks appear to tick for every tick of moving ones?''}

The method used here is exactly the same as was outlined in \RefSubsec{Subsec:Time dilation as seen by stationary observers}. Consider that some clock in the moving chain, $k$, is next to a clock in the stationary chain, $z$, when the stationary clock has a proper clock reading $\nu_z\stch$. At some later point in time, the moving chain has moved such that some other clock in the moving chain, $l$, is next to clock $z$ at the moment that clock $z$ advances its proper clock reading forwards one tick to $\nu_z\stch+1$. The difference in time, $\Delta{t}$, that it takes for clock $z$ to tick once, $\Delta{\nu_z\stch}=1$ can be obtained using \RefEq{Eq:Difference in time as a function of difference in proper clock reading} (with $\beta=0$ because we are considering the stationary clock). The time taken for a stationary sound clock to advance its proper clock reading by one tick is found to be $2L/\SoS$ (as expected from \RefEq{Eq:Period of stationary sound clock}). The distance that the \textit{moving} chain has travelled in this time is given by
\begin{equation}
x\moch = v\Delta{t} = 2L\beta.
\end{equation}
Sound clocks within the moving chain are separated by a distance of $L/\gamma$. The distance that the moving sound clock chain has travelled, $x\moch$, is equal to some multiple, $b$, of its own sound clocks' separation length,
\begin{equation}
b\dfrac{L}{\gamma} = 2L\beta.
\end{equation}
From this, we can easily solve for $b$:
\begin{equation}
b = 2\gamma\beta
\label{Eq:Moving experiment Number of clocks in stationary chain that the moving chain has moved over in one stationary tick}
\end{equation}
Note that the value of $b$ determined here is exactly the same as $b$ in \RefEq{Eq:Stationary experiment Number of clocks in stationary chain that the moving chain has moved over in one moving tick}.

The moving sound clock chain has travelled a distance of $b$ multiples of its own spacing arm length in the time that it has taken the stationary sound clock, $z$, to tick once. Therefore, clock $l=k-b$ is next to the stationary clock, $z$, when it advances its time forward by one tick. The difference between the proper clock readings of clock $k$ at some time $t_0$ and clock $l$ at some later time $t_1 = t_0 + 2L/\SoS$ can be obtained from \RefEq{Eq:Difference in proper clock reading as a function of difference in time}:
\begin{equation}
\Delta{\nu\moch}\left(\Delta{t}\right)=\nu_l\moch(t_1)-\nu_k\moch(t_0)=\gamma.
\label{Eq:Moving observers measure time dilation}
\end{equation}
Where, again, the superscript M indicates that these are quantities pertaining to the moving chain. Even though the chain of moving sound clocks actually ticks less frequently than the chain of stationary sound clocks, observers travelling along with the moving chain believe that the stationary chain ticks less frequently than their own due to their incorrect belief that they are at rest, which results in their clocks being asynchronous.

\subsection{Length contraction as seen by moving observers}
\label{Subsec:Length contraction as seen by moving observers}

The observers in the moving chain now decide to focus on more than just the first sound clock of the stationary chain in order to perform their next experiment:

\textit{``How are clocks in the stationary chain spaced, relative to clocks in the moving chain?’’}

We take the same approach as with the stationary chain's experiment, except with the roles of the stationary and moving chain reversed: which two clocks in the moving chain are two adjacent clocks in the stationary chain simultaneously next to (where simultaneity is defined as equal proper clock readings)?

\begin{figure}[tb]
	\centering
	\includegraphics[width=\linewidth]{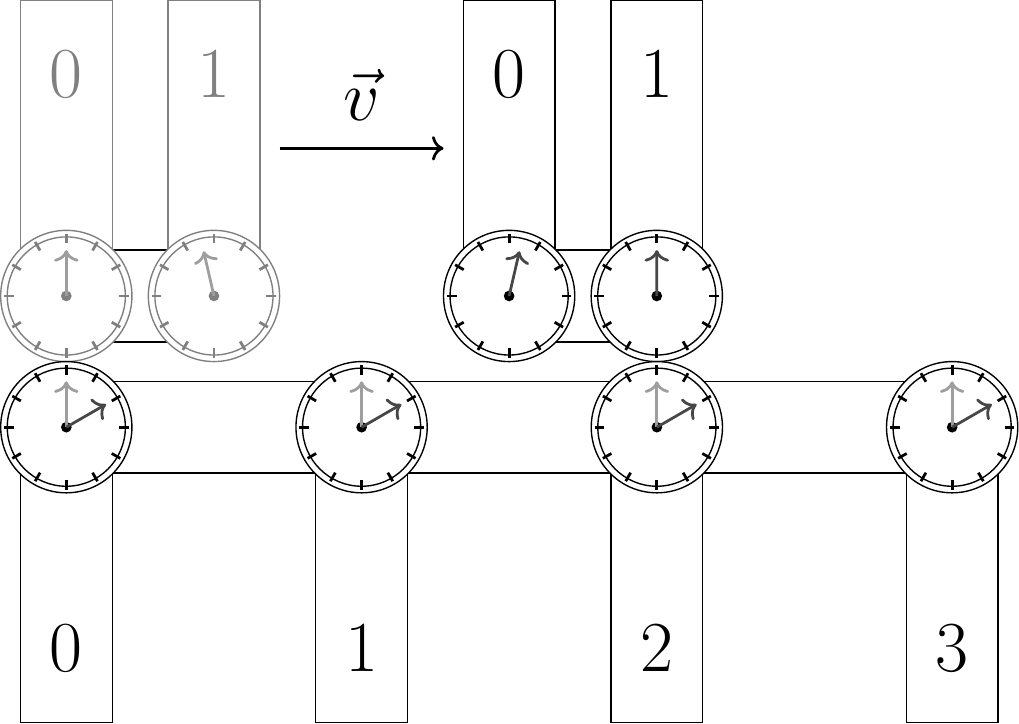}
	\caption[The observers in the moving chain determine what they believe to be the separation of sound clocks in the stationary chain by determining how many sound clocks are between two of their own sound clocks at an instant in time (as they understand it). The moving chain has a fractional velocity of $\beta=\sqrt{3}/2$, with respect to sound, corresponding to a Lorentz factor of $\gamma=2$. At this velocity, adjacent clocks in the moving chain only have the same clock reading when separated by a period of time that corresponds to having travelled a distance of $\gamma{L} = 2L$ in the laboratory. In this example, clocks $k$ and $l$ in the moving chain are clocks $0$ and $1$, respectively.]{The observers in the moving chain determine what they believe to be the separation of sound clocks in the stationary chain by determining how many sound clocks are between two of their own sound clocks at an instant in time (as they understand it). The moving chain has a fractional velocity of $\beta=\sqrt{3}/2$, with respect to sound, corresponding to a Lorentz factor of $\gamma=2$. At this velocity, adjacent clocks in the moving chain only have the same clock reading when separated by a period of time that corresponds to having travelled a distance of $\gamma{L} = 2L$ in the laboratory. In this example, clocks $k$ and $l$ in the moving chain are clocks $0$ and $1$, respectively.}
	\label{img:Sonic length contraction chain A}
\end{figure}
The method used here is exactly the same as was outlined in \RefSubsec{Subsec:Length contraction as seen by stationary observers}. Consider the scenario in which at some time, $t_0$, some clock in the moving chain, $k$, has proper clock reading $\nu_k\moch$ while next to a clock in the stationary chain. After what period of time does another clock $l$, that is $b$ spacing arms away ($l=k+b$), have the same proper clock reading as clock $k$ ($\nu_l\moch=\nu_k\moch$)? With use of \RefEq{Eq:Difference in time as a function of difference in proper clock reading} we can find the difference in time, $\Delta{t}$, for $\Delta{\nu\moch}=0$ for clocks $k$ and $l=k+b$:
\begin{equation}
\Delta{t}\left(\Delta{\nu}\moch\right) = t(\nu_l\moch) - t(\nu_k\moch) = b\dfrac{L\gamma\beta}{\SoS}.
\end{equation}
In this time the moving chain has moved a total distance of,
\begin{equation}
x\moch = v\Delta{t} = bL\gamma\beta^2.
\label{Eq:Distance travelled by moving chain for same proper clock reading}
\end{equation}
Clocks in the moving chain are separated by a distance of $L/\gamma$, so clock $l=k+b$ is $b$ multiples of $L/\gamma$ away from clock $k$. Clock $l$ has also travelled a total distance of $x\moch$ (from \RefEq{Eq:Distance travelled by moving chain for same proper clock reading}) after the moment in time when clock $k$ made its measurement, so the location of clock $l$ at time $t_1$ as compared the location of clock $k$ at time $t_0$ is given by
\begin{equation}
x_l\moch(t_1) = b\dfrac{L}{\gamma} + bL\gamma\beta^2. 
\end{equation}
This distance corresponds to some multiple, $b^\prime$, of the \textit{stationary} sound clocks' spacing arm length. Equating $b^{\prime}L$ with $x_l\moch(t_1)$ yields
\begin{equation}
b^{\prime} = {\gamma}b.
\label{Eq:Moving chain measures number of stationary clocks}
\end{equation}
Clocks $k$ and $l=k+b$ in the moving chain register the same proper clock reading at two different instants in time. At these instants in time, clock $k$ is situated over some clock in the stationary chain, and clock $l=k+b$ is situated over some other clock that is ${\gamma b}$ spacing arms away from the clock that $k$ was situated over, as can be seen in \RefFig{img:Sonic length contraction chain A}, for $b=1$. Directly adjacent neighbours in the moving chain (\textit{i.e.} $b=1$ in \RefEq{Eq:Moving chain measures number of stationary clocks}) will (incorrectly) conclude that there are exactly $\gamma$ clock arms simultaneously parallel to their single clock arm, a result that arises due to the asynchronicity of their clocks.

\section{Sonic relativity}
\label{Sonic relativity}

We have now seen what look like relativistic effects from the perspective of internal observers: time dilatation appears to be given by \RefEq{Eq:Stationary observers measure time dilation} and \RefEq{Eq:Moving observers measure time dilation}, while apparent length contraction appears to be described by equations \RefEq{Eq:Stationary chain measures number of moving clocks} and \RefEq{Eq:Moving chain measures number of stationary clocks}. However, none of these relationships explicitly deal with lengths or times. Ticks and numbers of clocks are both unitless, and furthermore, while $\gamma$ as it appears in these equations is a quantity that we can calculate in the laboratory, internal observers as we have currently described them can only \textit{measure} it (by counting how many clocks they pass in a given period of time or by comparing clock readings on their own chain to another chain). We shall now cast all previous formulae in terms of quantities that internal observers themselves can measure and demonstrate the appearance of sonic relativity. Let us start with some defined quantities based on beliefs held by internal observers.

Assume that observers operationally define a unit of length called an `\textit{arm}' and a unit of time called a `\textit{tic}'. A \textit{timing arm} of length $L$ (as measured in the laboratory) is operationally defined by internal observers to be 1~arm long, and 1~tic is operationally defined to be the time it takes for a sound clock with a timing arm of length 1~arm to advance its clock reading forward once. All observers in inertial motion believe themselves to be at rest, and so they believe 1~tic of time to be the time it takes for a sound pulse to travel 2 arm (to the end of the timing arm and back again). The speed of sound to any internal observer is then \textit{defined} to be
\begin{equation}
\tilde{\SoS}\coloneqq 2 \dfrac{\mathrm{arm}}{\mathrm{tic}}.
\label{Eq:Internal observers speed of sound definition}
\end{equation}
With the belief that they are at rest, all observers co-moving with a chain of sound clocks that are calibrated and synchronised as per the procedures outlined in \RefSubsec{Subsec: Calibrating clock separation} and \RefSubsec{Subsec: Synchronisation of clocks} believe that their \textit{spacing arms} are exactly 1~arm in length each, as it takes exactly 1~tic of time for the echo of a sound pulse propagated between adjacent clocks in a chain to return. This requirement itself formed the basis of the calibration procedure as outlined in \RefSubsec{Subsec: Calibrating clock separation}. With these definitions, we can determine at what velocity~--~in units of arm/tic~--~observers in a given chain believe another chain to be travelling.

In 1~tic of time for a stationary clock, $2\gamma\beta$ clock arms in the moving chain pass by (as per \RefEq{Eq:Moving experiment Number of clocks in stationary chain that the moving chain has moved over in one stationary tick}). Recalling from \RefEq{Eq:Stationary chain measures number of moving clocks} that stationary observers believe that $\gamma$ spacing arms in the moving chain are simultaneously next to one of their own, observers on the stationary chain come to conclude that clocks in the moving chain are separated by $\gamma^{-1}$ arm per clock. Stationary observers therefore (correctly) believe that $2\gamma\beta$ clocks in the moving chain have a length of $2\beta$ arm. These $2\gamma\beta$ sound clocks of length $2\beta$ arm take 1~tic to pass by. Thus, the perceived velocity of the moving chain, $\tilde{v}$, in units of arm/tic (up to a sign) is given by
\begin{equation}
\left|\tilde{v}\right| = \dfrac{2\beta~\mathrm{arm}}{1~\mathrm{tic}} = 2\dfrac{v}{\SoS}\dfrac{\mathrm{arm}}{\mathrm{tic}} = \dfrac{\tilde{\SoS}}{\SoS}v = \tilde{\SoS}\beta.
\end{equation}
The perceived fractional velocity, $\tilde{\beta}\coloneqq\tilde{v}/\tilde{\SoS}$, of the moving chain with respect to sound using only measurements of quantities available to internal observers is related to the \textit{actual} fractional velocity with respect to sound by
\begin{equation}
\left|\tilde{\beta}\right| = \beta.
\end{equation}

Moving observes determine the same relationships. A moving clock passes by $2\gamma\beta$ stationary clocks in 1~tic of time (as per \RefEq{Eq:Stationary experiment Number of clocks in stationary chain that the moving chain has moved over in one moving tick}), and from \RefEq{Eq:Moving chain measures number of stationary clocks} moving observers deduce that $\gamma$ spacing arms in the stationary chain lay simultaneously parallel to one of their own (\textit{i.e.} moving clocks appear to be separated by $\gamma^{-1}$ arm per clock). Because the moving observers think that they are the ones who are stationary and that the stationary observers are moving, they believe that $2\gamma\beta$ clocks have passed by them, with a total length of $2\beta$ arm in a time of 1~tic. This leads to the (believed) velocity of the moving chain, $\tilde{v}$, in units of arm/tic (up to a sign) of
\begin{equation}
\left|\tilde{v}\right| = \dfrac{2\beta~\textrm{arm}}{1~\textrm{tic}} = 2\dfrac{v}{\SoS}\dfrac{\textrm{arm}}{\textrm{tic}} = \dfrac{\tilde{\SoS}}{\SoS}v = \tilde{\SoS}\beta.
\end{equation}
The believed fractional velocity, $\tilde{\beta}\coloneqq\tilde{v}/\tilde{\SoS}$, of the stationary chain with respect to sound using only measurements of quantities available to internal observers is given by
\begin{equation}
\left|\tilde{\beta}\right| = \beta.
\end{equation}

Both stationary and moving observers believe that the other chain of sound clocks is moving with a fractional velocity with respect to sound that is equal in magnitude to the value of the moving chain's fractional velocity with respect to sound in the laboratory frame. Utilising an agreed-upon coordinate system, observers in the moving chain will report the value of $\tilde{\beta}$ that they measure to be different than that as measured by stationary observers by a sign. Both stationary and moving observers define
\begin{equation}
\tilde{\gamma}\coloneqq\dfrac{1}{\sqrt{1-\tilde{\beta}^2}}.
\end{equation}

A given observer measures length by counting the number of spacing arms between simultaneous measurements of the endpoints of an object: 1~arm is exactly the length of one spacing arm. We define $\tilde{\ell}^\prime$~to be the length, as measured by a stationary observer, of an object whose length is measured to be~$\tilde\ell$ by an observer at rest with respect to the object. Both stationary \textit{and} moving observers believe that ${\gamma}b$ clocks in the other chain lay simultaneously next to $b$ of their own, as per \RefEq{Eq:Stationary chain measures number of moving clocks} and \RefEq{Eq:Moving chain measures number of stationary clocks}, respectively. Thus, observers in each frame will state that the total length spanned by some number of their own spacing arms is equal to the length spanned by ${\gamma}$ times as many spacing arms of the other chain, leading to the relationship
\begin{eqnarray}
\tilde{\ell}^\prime = \dfrac{\tilde{\ell}}{\tilde{\gamma}},
\end{eqnarray}

where $\gamma$ has been replaced by the internal-observer-defined $\tilde{\gamma}$.

A given observer measures elapsed time by counting ticks of their own clock: 1~tic is exactly the time it takes for one's own clock to tick once. We define $\tilde{\tau}^\prime$~to be the duration, as measured by a stationary observer, of a localised process whose duration is measured to be~$\tilde\tau$ by an observer at rest with respect to the process. As per \RefEq{Eq:Stationary observers measure time dilation} and \RefEq{Eq:Moving observers measure time dilation}, both stationary \textit{and} moving observers believe their own clock readings to have advanced $\tilde{\gamma}$ times in the time it takes a given clock in the other chain to advance its clock reading once. Therefore, both stationary and moving observers believe that a clock in the other chain takes $\tilde{\gamma}$ as much time to tick once as their own clock does, leading to the relationship
\begin{equation}
\tilde{\tau}^\prime = \tilde{\gamma}\tilde{\tau},
\end{equation}
where, as before, the internal-observer-defined Lorentz factor $\tilde{\gamma}$ is used due to its equality with $\gamma$.

\section{Discussion}
\label{Sec:Discussion}

Through an operational approach, it has been shown that it is possible for a certain class of inertial observers to deduce the existence of two key phenomena from special relativity~--~length contraction and time dilation~--~in condensed-matter systems for which the speed of sound plays an analogous role to the speed of light within our universe. The observers we discuss are significantly restricted in their ability to assign temporal and spatial values to events. They are only able to claim when events occur relative to their own clock, can only make claims about events that are sufficiently local such that the time it takes for the signal to reach them is negligible, and must confer with one another after taking local measurements to come to an understanding of the events that transpired.

The observers travelling at a constant velocity have no way to tell if they are stationary or moving. With no way to tell who is in motion, this question would become a philosophical one for internal observers, and some internal observers may even come to the same conclusions that we in our universe have: that all motion is relative. We have seen that when internal observers assume the former state of motion~--~that they are indeed stationary~--~then those who are in motion incorrectly set the separation of their clocks, a process that was achieved by fulfilling the requirement that sound simultaneously emitted along two objects of equal length will result in the simultaneous reception of echoes. This simultaneity condition results not only in the incorrect separation of clocks within a moving chain; it also results in the asynchronicity of those clocks.

In constructing their chains of sound clocks in such a way that local simultaneity conditions hold, observers who are stationary see the moving chain to be length contracted exactly as one would expect from a na\"{i}ve application of relativistic formulae, with $\SoS$ being the speed of sound instead of the speed of light. The clocks within the moving chain also appear to be time dilated as would be expected from special relativity. This is due to the use of sound pulses to advance clock readings, and moving clocks increase the path length, thus increasing the time it takes for a sound pulse to return to the clock mechanism by exactly the Lorentz factor again.

More curiously, the observers in moving chains also witness stationary sound clock chains as being length contracted: this is not actually the case and is again a result of making use of simultaneity arguments. Moving observers \textit{think} that the clocks within their own chain are synchronous, and thus when they wish to know what happened at some distant clock simultaneous with their own clock, they ask the observer located at that distant clock to provide them with information recorded when the distant clock's reading was the same as their own. These clocks are not actually synchronous, however, so the observers in the moving chain are actually comparing information from two separate instances in time. This happens to work out in exactly the right way to make the observations of moving observers and stationary observers symmetric: moving observers also perceive stationary sound clock chains to be length contracted and time dilated exactly as one would expect from a na\"{i}ve application of relativistic formulae, with $\SoS$ being the speed of sound instead of the speed of light.

We see that the `in-universe' experience~--~the internal observers' description of their universe~--~can be described by the mathematical formalism of special relativity provided that such observers believe the postulates of relativity, a conclusion that they would reasonably come to when given no ability to detect their aether. It is merely a misunderstanding of the Newtonian mechanics at play that results in the appearance of relativistic effects to these internal observers. Given the ability to detect their own state of motion relative to their aether, moving observers would quickly come to understand that they have incorrectly separated and calibrated their clocks, and fixing this problem would result in the disappearance of the apparent relativistic effects that are witnessed by moving observers.

We have intentionally remained within the realm of discussing devices that are operationally controlled by observers, and the relativity within the system described appears as a result of the belief that observers have about their state of motion. Nevertheless, it is worth noting that if we had access to devices that were built from quasiparticles made up from the medium itself, then sonic relativity would emerge naturally. As described by Barcel\'{o} and Jannes~\cite{Barcelo2008}, a device constructed from these quasiparticles (such as a sound clock chain) would shrink naturally as $v$ approaches $\SoS$, just as physical objects held together by the electromagnetic force do when travelling close to the speed of light~\cite{Bell1976}. This would entirely remove the role of the observers in tuning the separation of neighbouring clocks, and therefore the belief held by the observers on their state of motion would become inconsequential.

Furthermore, while we have restricted our analysis to chains of sound clocks that must keep their timing and spacing arms perpendicular, and that are only able to move in the direction perpendicular to their timing arms, devices built from quasiparticles would not be limited in these ways. If the sound clock chains that we describe here were permitted to travel with a velocity possessing a non-zero component in the direction parallel with the timing arms, then any change in the direction of motion mid-journey to include such a component of velocity would lead to the asynchronous return of echoes in the timing and spacing arms. Additionally, if these devices could alter the angle between their arms, this would again lead to the asynchronous return of sound pulses in the timing and spacing arms, and if the angle between the arms was closed to $0^\circ$, observers could quite easily determine that the timing and spacing arms are different lengths. These effects would be naturally taken care of in quasiparticle devices, however: the entire device's dimensions would change accordingly with its velocity, conspiring to make the effects of length contraction impossible to detect to the observers located on the chain.

\section{Conclusion}
\label{Sec:Conclusion}

It is perhaps not too surprising that stationary observers in these systems infer that moving observers undergo relativistic time dilation and length contraction, but given the presence of a preferred reference frame it may not be immediately obvious that moving observers should see stationary observers subject to these same effects (although if one has a sufficient understanding of the history of special relativity, especially in the context of aether theory, then this may not actually be too surprising). Given that relativity \textit{is} seen in both directions just as in our universe, it can be seen that the existence of a preferred reference frame is not immediately prohibitive in the emergence of a self-consistent description of relativity by internal observers in analogue gravity systems. While some aspects of relativity can be made to appear, it is not clear to what extent relativistic physics can be made to manifest in such systems.

If the role of observers in analogue gravity systems is taken more seriously, investigating the observations made by such observers might give us some insight into how many of the phenomena described by general relativity can be seen to arise in analogue gravity models in a self-consistent manner. Are there any analogue gravity models that appear to possess all of the phenomena of general relativity (in analogous forms) to internal observers? If this is not the case, then why not? Why should some of the phenomena of relativity emerge in a self-consistent manner in such models, but not others? Considering the experience of observers may help to answer these questions.

\section*{Acknowledgements}

We thank Magdalena Zych, Jason Pye, Robert Mann, Fabio Costa, and Ian Durham for inspiring discussions. We also thank the anonymous referee for suggesting additional references that will be of use to us in future work. This work was supported by the U.S.\ Air Force Research Laboratory (AFRL) Asian Office of Aerospace Research and Development (AOARD) under grant number FA2386-16-1-4020 and by the Australian Research Council Centre of Excellence for Quantum Computation and Communication Technology (Project No.~CE170100012).

\bibliography{PhDBib}

\end{document}